\pdfoutput=1
\documentclass{JINST}

\title{Front-end Multiplexing - applied to SQUID multiplexing : Athena X-IFU and QUBIC experiments}

\author{Damien PRELE \\
\llap APC,\\
AstroParticule et Cosmologie, Universit\'e Paris Diderot, CNRS/IN2P3, CEA/Irfu, Obs. de Paris, Sorbonne Paris Cit\'e, 10, rue Alice Domon et L\'eonie Duquet, 75205, Paris Cedex 13, France\\
  E-mail: \email{prele@apc.in2p3.fr}
 }

\abstract{As we have seen for digital camera market and a sensor resolution increasing to "megapixels", all the scientific and high-tech imagers (whatever the wave length - from radio to X-ray range) tends also to always increases the pixels number. So the constraints on front-end signals transmission increase too. An almost unavoidable solution to simplify integration of large arrays of pixels is front-end multiplexing. Moreover, "simple" and "efficient" techniques allow integration of read-out multiplexers in the focal plane itself. For instance, CCD (Charge Coupled Device) technology has boost number of pixels in digital camera. Indeed, this is exactly a planar technology which integrates both the sensors and a front-end multiplexed readout. In this context, front-end multiplexing techniques will be discussed for a better understanding of their advantages and their limits. Finally, the cases of astronomical instruments in the millimeter and in the X-ray ranges using SQUID (Superconducting QUantum Interference Device) will be described.}


\begin{document}

\section{Introduction} 

\begin{figure}[h]
	\begin{center}
		\includegraphics[width=0.7\linewidth, keepaspectratio]{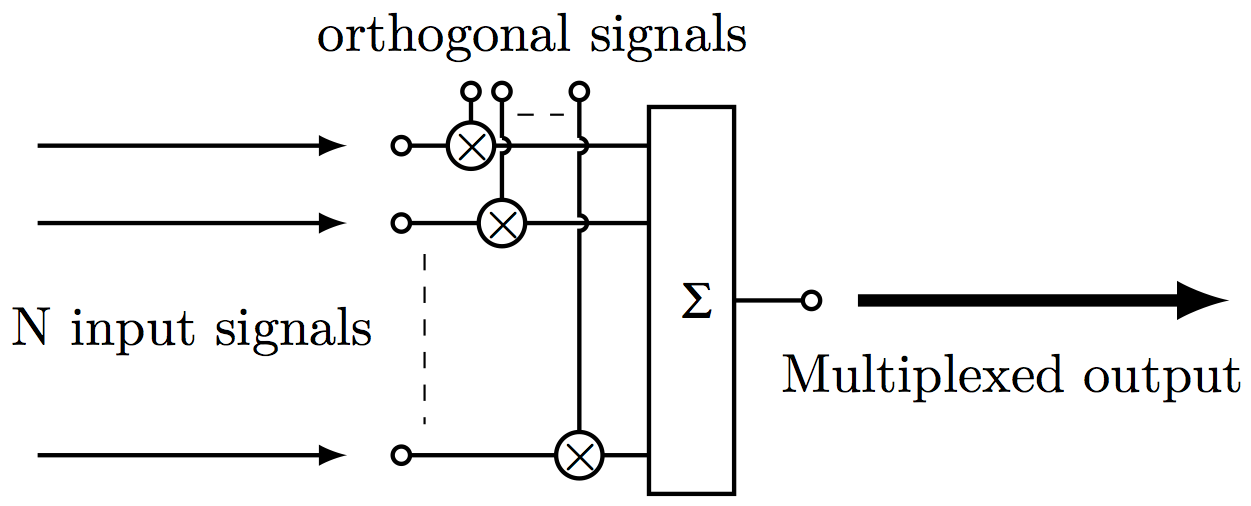}
	\end{center}
	\caption{Multiplexing as a summation of orthogonal carrier signals modulated by input signals.}
	\label{Fig1}
\end{figure}

Multiplexing allows to transmit several data/signal (informations) by using a single, high "capacity", channel. As it is represented in figure~\ref{Fig1}, multiplexing can be split into two stages : {\bf Modulation} by orthogonal carrier signals and {\bf Summation} of the modulated carriers by the input signals.\\

\begin{figure}[h!]
	\begin{center}
		\includegraphics[width=.95\linewidth, keepaspectratio]{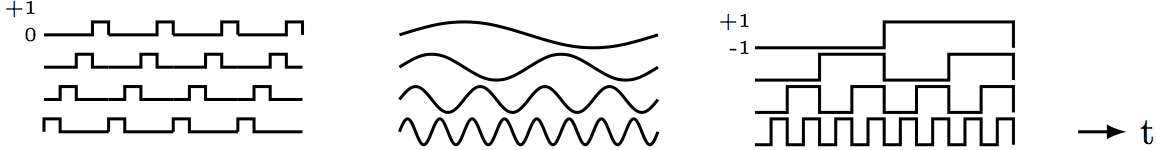}
	\end{center}
	\caption{Shape of the carriers - orthogonal signals - used for a 4 to 1 time, frequency and coded division multiplexing.}
	\label{Fig2}
\end{figure}

Signals used as carriers for modulation should be chosen so that they associate a specific code to each input signal. So, this modulation could also be seen as a coding. The carriers used for the modulation/coding are said orthogonal, since they are used to restore each input signal independently. Thanks to that, the demultiplexer is able to recover each input signal without interferences. Square signals with small duty cycle ({\it boxcar}) and shifted in time, tones at different frequencies or Hadamard/Walsh code are typical carriers (Fig. \ref{Fig2}) used for modulation by input signals before summation, producing time (TDM), frequency (FDM) and coded (CDM) division 4 to 1 multiplexing, respectively.

\begin{figure}[h!]
	\begin{center}
		\includegraphics[width=.6\linewidth, keepaspectratio]{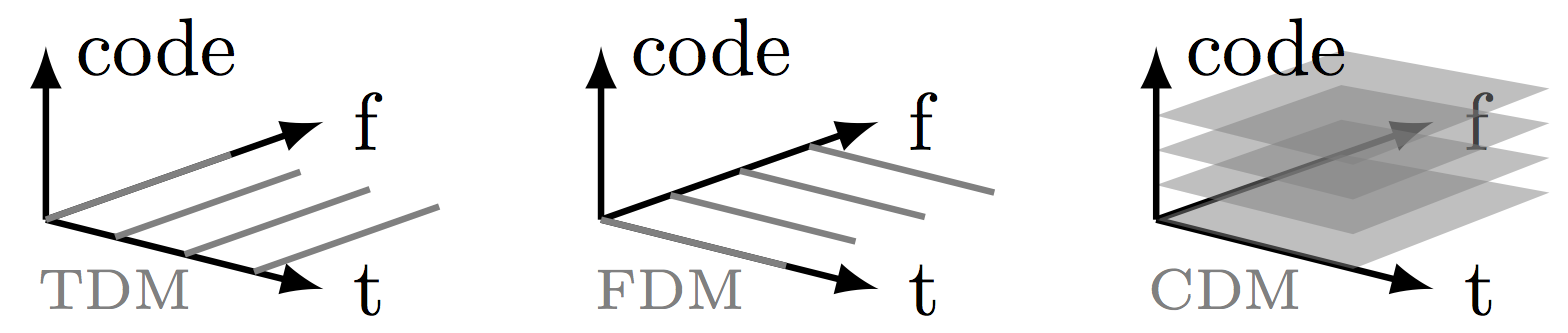}
	\end{center}
	\caption{Time, frequency and code : "3 dimensions" usable to do TDM, FDM and CDM.}
	\label{Fig3}
\end{figure}

Figure \ref{Fig2} shows typical shape of the carriers to sum without interference 4 signals into a single channel. The normalized amplitude of the square signals comes from the use of switches to do this kind of modulation. For coded division multiplexing, a quite more complicated switches bridge must be used to invert the read-out polarity ($\pm$ 1). Indeed, there is no OFF state in coded division multiplexing, neither in the frequency domain multiplexing. So, this consideration is often used to say that FDM and CDM have better performance due to full time reconstruction of the input signals. At the opposite, TDM samples the signal. But these modulated signals must be summed together (Fig. \ref{Fig1}) leading to a significant increase of the amplitude in the case of the FDM and CDM, while a time domain multiplexed signal keeps the same amplitude as the individual input signals. So many parameters must be taken into account to do a strict comparison between multiplexing techniques.\\
Figure \ref{Fig3} gives the time and frequency occupation of the TDM, FDM and CDM of these modulations. Code is represented as a third dimension even if this is not necessarily a physical dimension. Indeed, CDM is usually used to spread the spectrum of the multiplexed signal. But the code dimension is often a repartition both in time and in frequency. 

\paragraph{Performances :} To multiplex a signal, the readout system (multiplexer) must have better performance than to read-out a single pixel, otherwise degradation of the multiplexed signal occurs. The multiplexer must have better bandwidth, dynamic range and/or noise performance than for the readout of one pixel. The required increase in performance for an $N$ to 1 multiplexer must be better by a factor of about $\sqrt{N}$ to few $N$. In any case, the larger the $N$ number of input signals, the larger the capacity of the multiplexed channel must be.\\

First of all, {\bf bandwidth} must be increased. Whether for FDM, TDM or for CDM, the bandwidth needed is increased by more than a factor of $N$ (where $N$ is the number of multiplexed signal) because of the modulation. {\bf FDM} transposes each input $B$ band around a specific $c_1$, $c_2$, ... carrier spaced by $2B$ at least (with dual side-band modulation), so the needed bandwidth is at least equal to $2NB$. {\bf TDM} does samples of each input signal at a specific time (each $T_s$). The sampling frequency $f_s=\frac{1}{T_s}$ must be larger than two times the input bandwidth $B$ (Nyquist-Shannon theorem). Moreover, the small duty cycle (inversely proportional to $N$) square signal used for this sampling has a spectrum extended up to infinity (cardinal sine : $sinc$) and which can not be limited below $Nf_s$, so $2NB$ (main $sinc$ lobe).\\

{\bf Dynamic range} and signal to noise ratio (SNR) are also affected by the multiplexing. It is evident that because multiplexing is a summation of modulated signals, the amplitude of the multiplexed signal could be increased by the summation. And it is also obvious that the larger is the number $N$ of multiplexed inputs, the larger the amplitude of the result of the summation can be. In fact, it is not the case for TDM because inputs are sampled sequentially, only one at one time, not at the same time. As a result, the TDM signal has the same dynamic range than input signals. However, for FDM and CDM, the amplitude of the multiplexed signal can be as large as $N$ times the amplitude of input signals depending on the relative phase between each carrier (or other orthogonal signals). In real devices, the dynamic range is limited both by saturation or other non linearities leading to a maximum operating range more or less linked to the power consumption of the read-out system. So, to the purpose of a strict comparison of multiplexing techniques, it is important to consider a limited dynamic range of the channel. Thus, to counteract this amplitude increase, it is usually needed to attenuate the amplitude (or, at least, reduce the gain) of the frequency or coded multiplexed signal, as compare to TDM. For a same channel noise contribution, the SNR is then automatically reduced. 

\section{Multiplexing} 

From now on, we will discuss exclusively time and frequency division multiplexing. Indeed, this paper is a short review of front-end multiplexing techniques, so, coded division multiplexing, even if widely used in telecommunication \cite{bib1}, is not often applied to the "front-end" sensors multiplexing. 


So, we discuss the example of 4 to 1 TDM and FDM to give an idea of the signal shape encountered in the multiplexer. We assume that input signals are 4 simple sine waves (Sig1, 2, 3 and 4 plot on top of figures~\ref{Fig6} and \ref{Fig9}): Unity in amplitude, roughly 100-200 Hz frequency and shifted in phase. Modulation is considered as ideal. 

\subsection{Time domain multiplexing}

\begin{figure}[h!]
		\includegraphics[scale = .25]{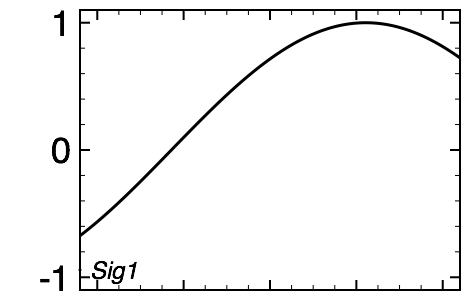}
		\includegraphics[scale = .25]{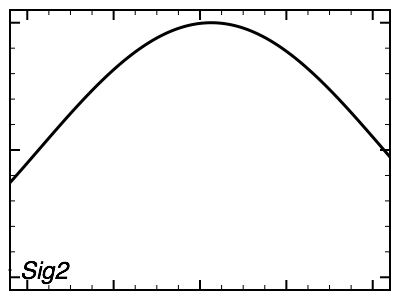}
		\includegraphics[scale = .25]{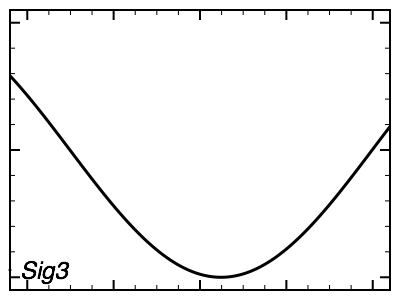}
		\includegraphics[scale = .25]{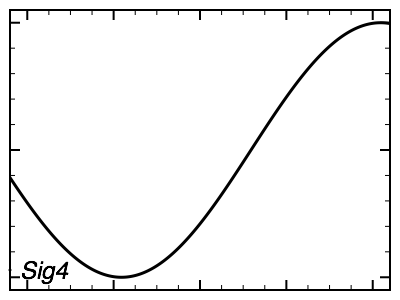}\\
		\includegraphics[scale = .25]{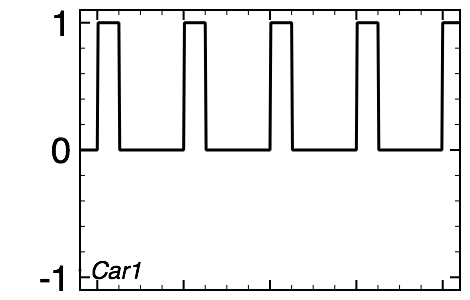}
		\includegraphics[scale = .25]{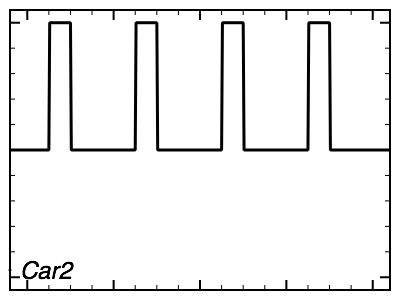}
		\includegraphics[scale = .25]{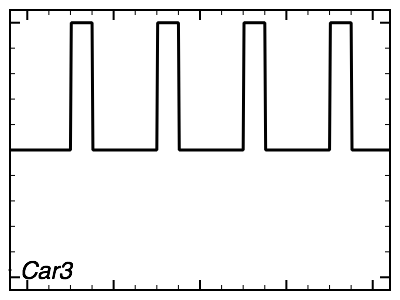}
		\includegraphics[scale = .25]{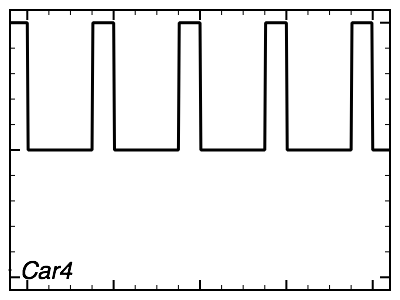}\\
		\includegraphics[scale = .25]{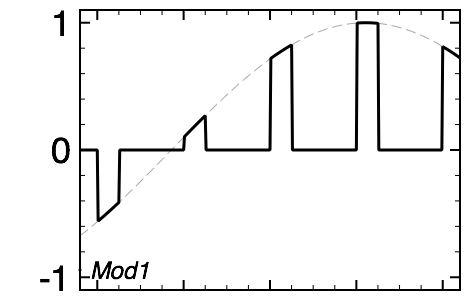}
		\includegraphics[scale = .25]{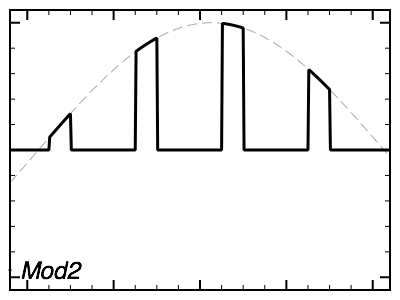}
		\includegraphics[scale = .25]{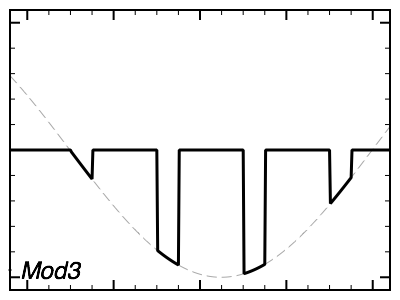}
		\includegraphics[scale = .25]{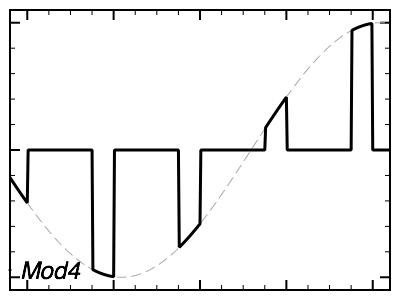}\\
		\includegraphics[scale = .25]{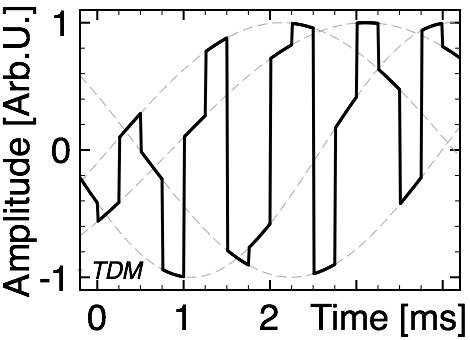}
		\includegraphics[scale = .25]{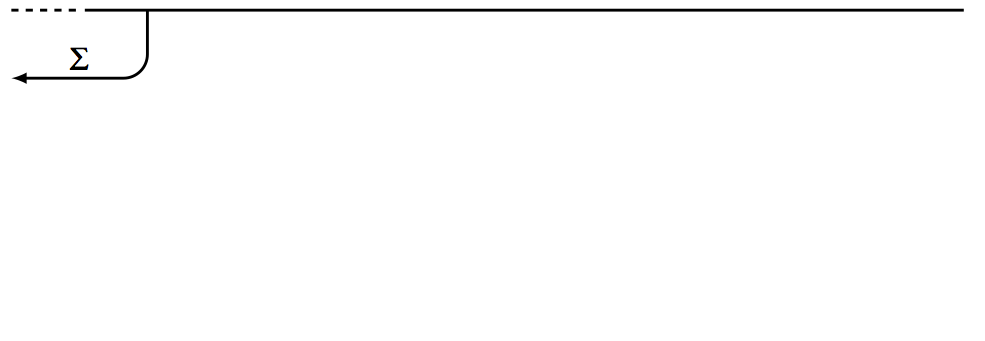}
	\caption{Example of 4 to 1 TDM. Input signals are simple sine waves (Sig\#). Carrier signals used for the modulation are perfect small duty cycle ($1/N$) square signals shifted in phase (Car\#). The modulations (Mod\#) appears clearly as a sampling of the input signals (dashed lines). Summation of samples shifted in time gives the shape of the time domain multiplexed signal (TDM).}
	\label{Fig6}
\end{figure}

Time domain multiplexing is a summation of limited duration time slot of each input signal (Fig.~\ref{Fig6}). The modulation of small duty cycle square signals (carriers : Car1, 2, 3 and 4) by input sine waves is similar to a sampling. 
The duty cycle of the carrier signals used in time domain multiplexing is equal to the "1" or "ON" duration divided by the sampling period $T_s=\frac{1}{f_s}$. Duty cycle is thus inversely proportional to the number $N$ of inputs (Eq. \ref{Eq1}).

 	\begin{equation}
   		Duty \; cycle = \frac{T_{ON}}{T_s} = \frac{1}{N}
		\label{Eq1} 
	\end{equation}
\vskip .2 cm

\begin{figure}[h!]
	\begin{center}
		\includegraphics[scale = .25]{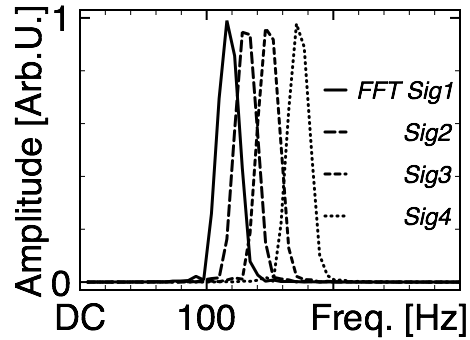}
		\includegraphics[scale = .25]{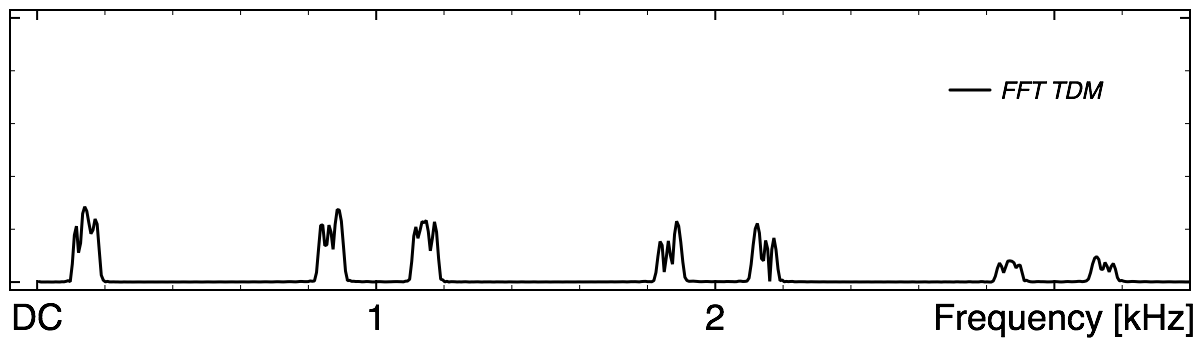}
	\end{center}
	\caption{Spectrum of the 4 input signals and of the multiplexed signal (same amplitude scale).}
	\label{Fig7}
\end{figure}

Dotted lines in the last plot of the figure~\ref{Fig6} allow to visualize that input signals are again visible in the waveform of the time domain multiplexed signal. Moreover, it is noticeable that multiplexed signals have the same amplitude (if no off-set) but occupy a higher band-width (due to fast transitions) than the input signals. Figure~\ref{Fig7} shows the spectrum of input signals and time domain multiplexed output signal. However, the spectrum of the time domain multiplexed signal does not give clear information on the input signals. Indeed, this spectrum mixes each modulation to the very near frequencies. 

\begin{figure}[h!]
		\includegraphics[scale = .25]{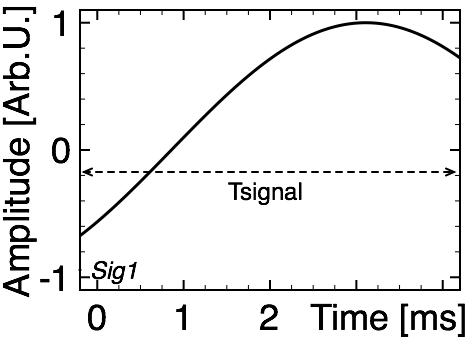}
		\includegraphics[scale = .25]{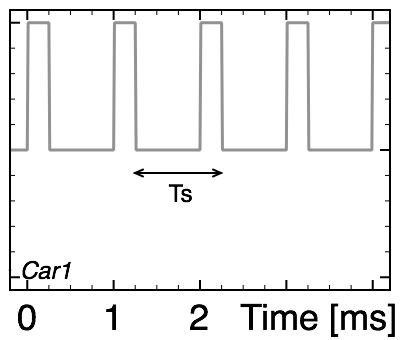}
		\includegraphics[scale = .25]{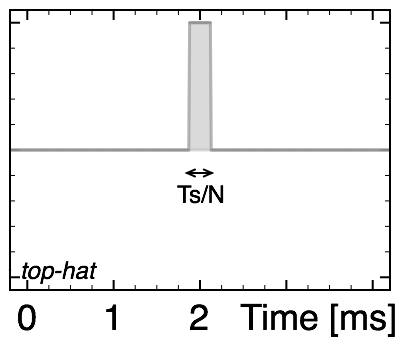}
		\includegraphics[scale = .25]{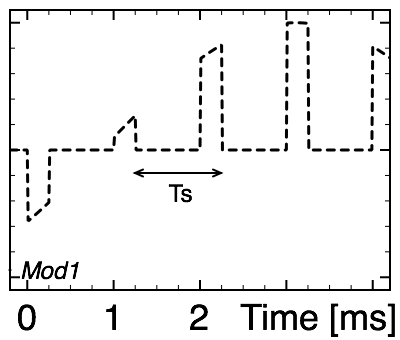}\\
		\includegraphics[scale = .25]{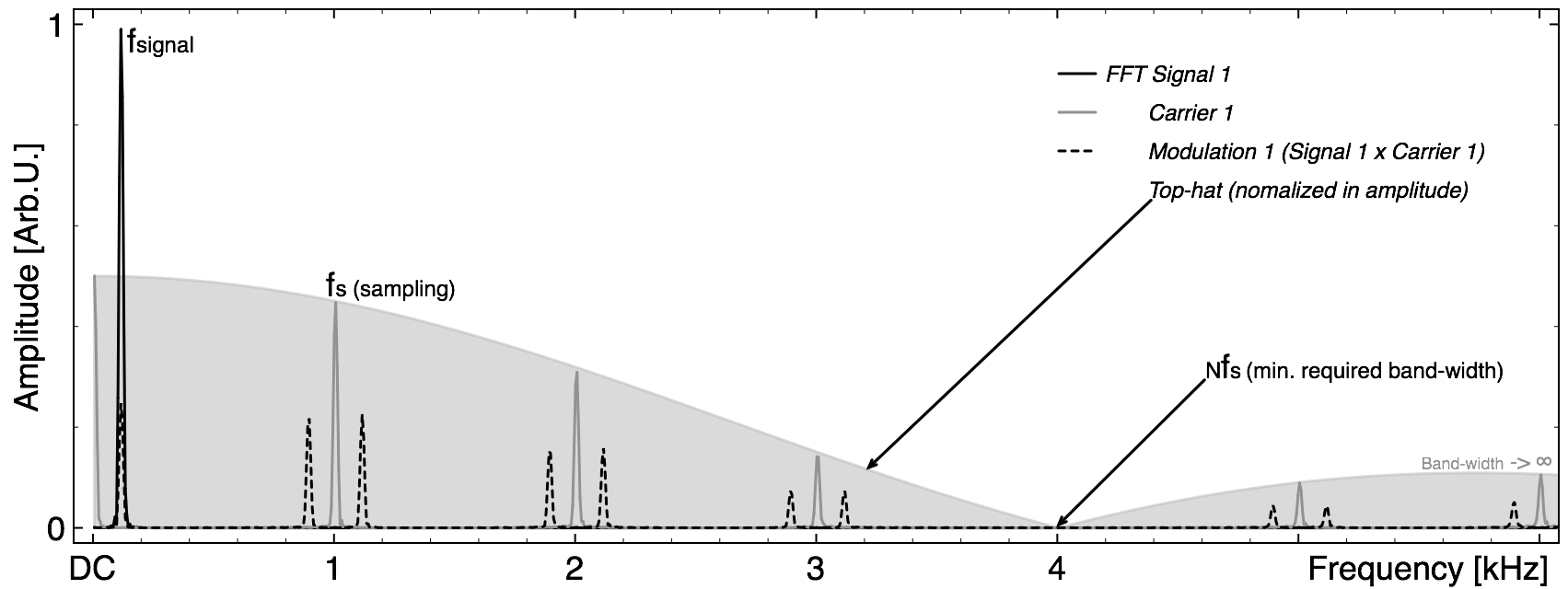}
	\caption{Frequency occupation of the time domain multiplexing : 
		{\bf{Input signal ($Sig1$) :}} Fourier transform of input sine-wave is a simple Dirac function centered on the frequency signal ; 
		{\bf{Carrier ($Car1$) :}} Spectrum of carrier signal - small duty cycle square signal - could be graphically obtained by multiplication of Dirac comb function by the spectrum of one {\it top-hat} function which composes the carrier ; 
		{\bf{Square function ({\it top-hat}) :}} The Fourier transform of a {\it top-hat} function is a cardinal sine (sinc) function ($\frac{A}{\pi f} \times \sin(\pi f T) = TA \times {\rm sinc}(\pi fT)$ with T the top-hat width and A the amplitude) and is represented in filled gray - normalized and in absolute amplitude value. 
		{\bf{Modulated signal ($Mod1$) :}} The amplitude modulation transposed input signal spectrum around each harmonic of the carrier signal. The modulation spectrum is represented in dotted-line in the figure and is, in fact, the spectrum of the multiplexed signal if the other inputs are $0$.
	}
	\label{Fig8}
\end{figure}

So, to understand the spectrum of time domain multiplexing, figure \ref{Fig8} gives link between wave-forms and spectrum of the modulation used in TDM.

It is noticeable that multiplexed spectrum has an infinite band-width following a cardinal sine. Nevertheless, a cardinal sine exhibits a main lobe which could correspond to a minimum required band-width as it is shown in figure \ref{Fig8}. This approximation is typically used to do practical sampling of the input signals\footnote{This practical band-width limitation introduces a cross-talk due to the slow transition of the multiplexed signal from one input to the next. This point is not developed in this review.}.
Finally, we can see that the {\it energy} corresponding to each input signal and carried by the multiplexed one is strongly reduced (in fact by a factor of $N$). This is the draw-back of the amplitude conservation of this multiplexing (Multiplexed signal dynamic range = input signal dynamic range).

\subsection{Frequency domain multiplexing}

\begin{figure}[h!]
		\includegraphics[scale = .25]{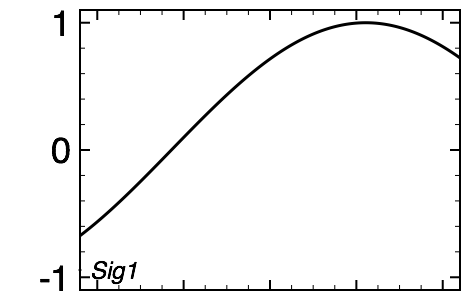}
		\includegraphics[scale = .25]{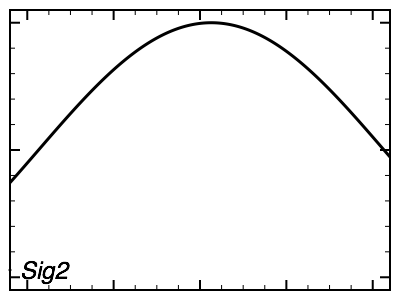}
		\includegraphics[scale = .25]{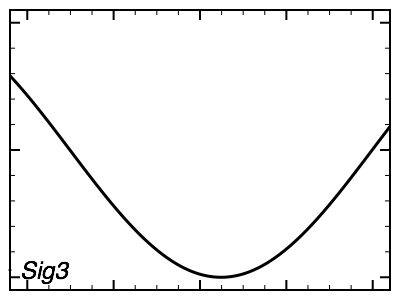}
		\includegraphics[scale = .25]{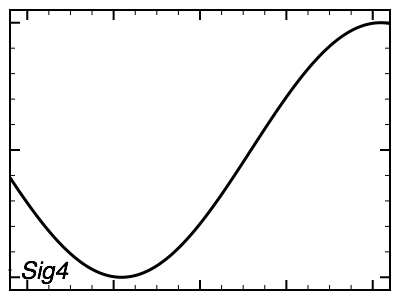}\\
		\includegraphics[scale = .25]{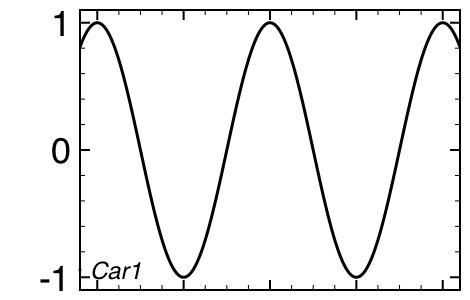}
		\includegraphics[scale = .25]{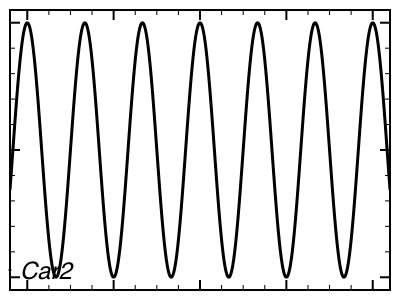}
		\includegraphics[scale = .25]{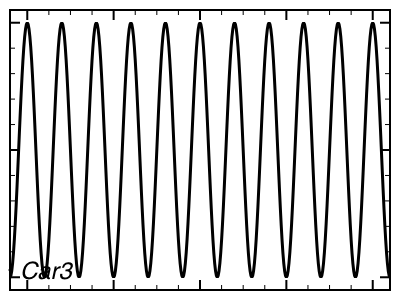}
		\includegraphics[scale = .25]{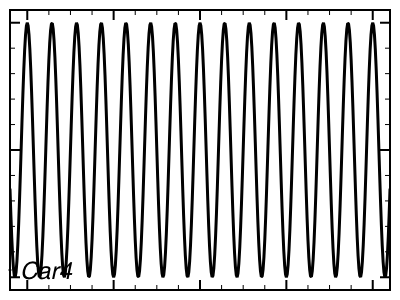}\\
		\includegraphics[scale = .25]{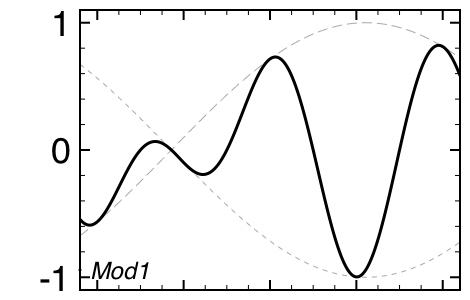}
		\includegraphics[scale = .25]{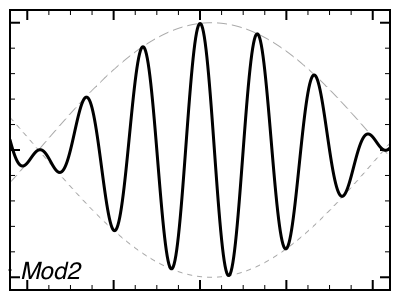}
		\includegraphics[scale = .25]{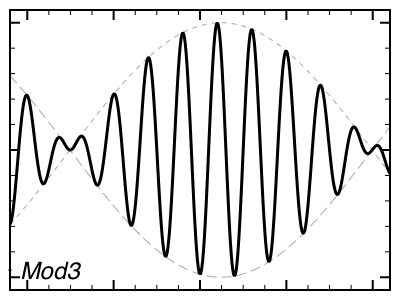}
		\includegraphics[scale = .25]{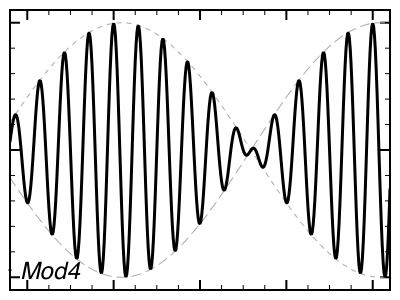}\\
		\includegraphics[scale = .25]{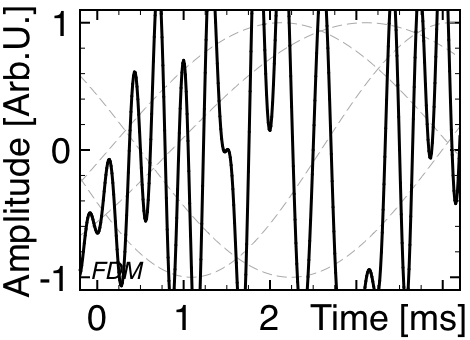}
		\includegraphics[scale = .25]{SumArrow.png}
	\caption{Example of 4 to 1 FDM. Input signals are simple sine waves (Sig\#). Carrier signals used for the modulation are sine waves at different frequencies (Car\#). Shape of input signals are again visible on the modulated signals (Mod\#) envelope (dashed lines). Summation of modulated signals gives the frequency domain multiplexed signal which is larger in amplitude and thus go beyond the frame.}
	\label{Fig9}
\end{figure}

Frequency domain multiplexing transposes each input signal around $N$ (4 here) sine wave carriers and sums the results of modulations. So, at the opposite of TDM where carriers are the same, only shifted in time, FDM requires a specific carrier frequency for each input. So $N$ carriers at frequencies spaced at least by $2$ times the signals band-width are needed. Figure \ref{Fig9} clearly shows sine-waves carriers at different frequencies (Car\#) and the results of the amplitude modulation (Mod\#) by input signals (Sig\#) for a 4 to 1 FDM. To minimize the multiplexed signal band-width, the first carrier frequency might be, at least, equal to the input band-width.

\begin{figure}[h!]
	\begin{center}
		\includegraphics[scale = .255]{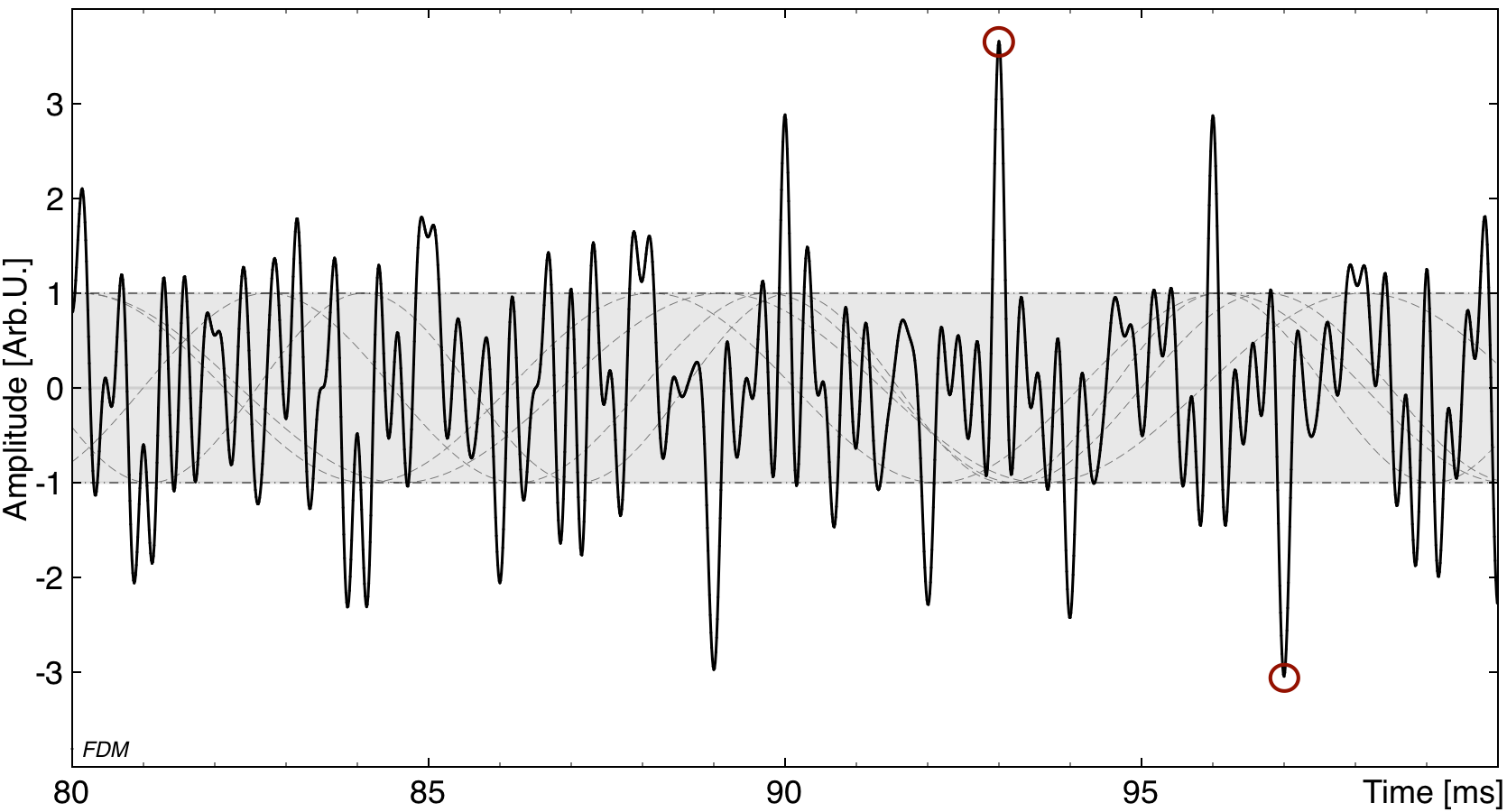}
	\end{center}
	\caption{Zoom-out on the multiplexed signal (FDM) exhibiting amplitude near 4 times input signals dynamic range. Small red circles tag the peak values. Gray zone point out input signals (dashed lines) dynamic range.}
	\label{Fig10}
\end{figure}

The band-width increase of the multiplexed signal as compared to the input signal band-width is obvious in the case of FDM. Indeed, a specific "frequency" channel is assigned to each $N$ input signal. So the band-width is necessarily increased by a factor proportional to $N$. Moreover, due to the double-side band modulation typically used, the band-width is in fact increase by a factor larger than $2N$. Finally, to avoid crosstalk between each adjacent carriers, the frequency difference between each carriers is greater than $2N$. So, in practice, the band-width increase is significantly worse than a factor $2N$.\\

In addition to the bandwidth widening, the frequency domain multiplexed signal is characterized by a dynamic amplitude also strongly increased, as shown in figure \ref{Fig10}. The figure clearly highlights the increase in the amplitude dynamic range as compared to the input signal (gray zone in the figure \ref{Fig10}). Two red circles point to peak values, $N$ times the signal amplitude range. 

\begin{figure}[h!]
	\begin{center}
		\includegraphics[scale = .25]{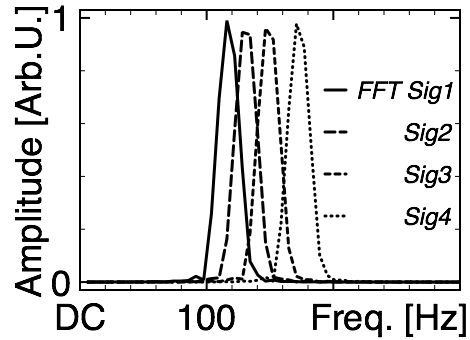}
		\includegraphics[scale = .25]{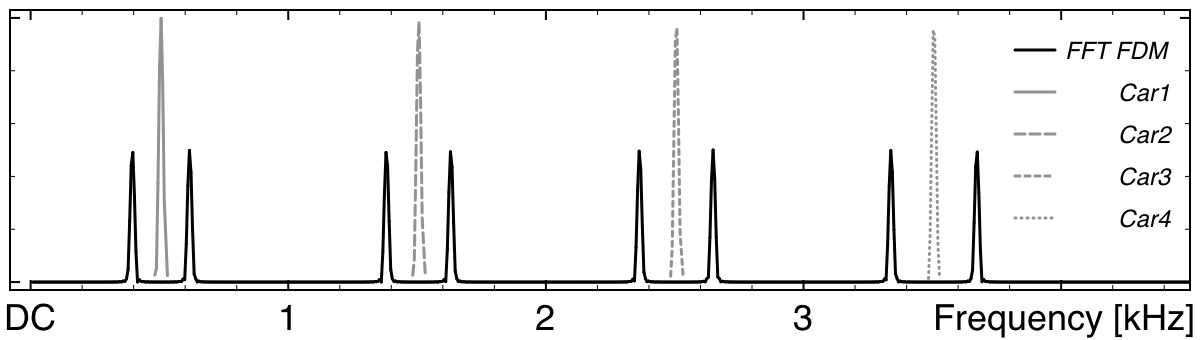}
	\end{center}
	\caption{Spectrum of the 4 input signals and of the multiplexed signal (solid black line - same amplitude scale). Carrier signals are plotted in gray/dashed lines.}
	\label{Fig11}
\end{figure}

It is hard to recognize input signals in the wave form of the multiplexed signal. Indeed, modulated signals are mixed together at any time (at the opposite of the TDM). Only the spectrum of the frequency multiplexed signal allows to trace each input signal. So, spectra of this multiplexing are plotted in figure~\ref{Fig11}. Spectrum of the carrier signals (gray/dashed line) locates the frequencies around which the input signals are transposed. Finally, solid line illustrates the sum of four standard double side band amplitude modulations.

\section{Applications} 

Front-end multiplexing corresponds to put a multiplexing system very near the detectors. This is particularly required if the number of sensors is large and/or if there is a strong constraint on the wiring. We have already mentioned the charge-coupled device (CCD) in the introduction as perfect example of front-end multiplexer. Indeed, CCD allows to do time domain multiplexing of a large array on the sensor wafer itself. In this part, we will briefly describe multiplexers (TDM and FDM) developed for modern astrophysic observations in the {\it mm} wave-length and {\it X-ray} range. We will start by some technological considerations to identify topologies and devices required for front-end multiplexing.

\subsection{Technological considerations}

\begin{figure}[h!]
	\begin{center}
		\includegraphics[scale = .24]{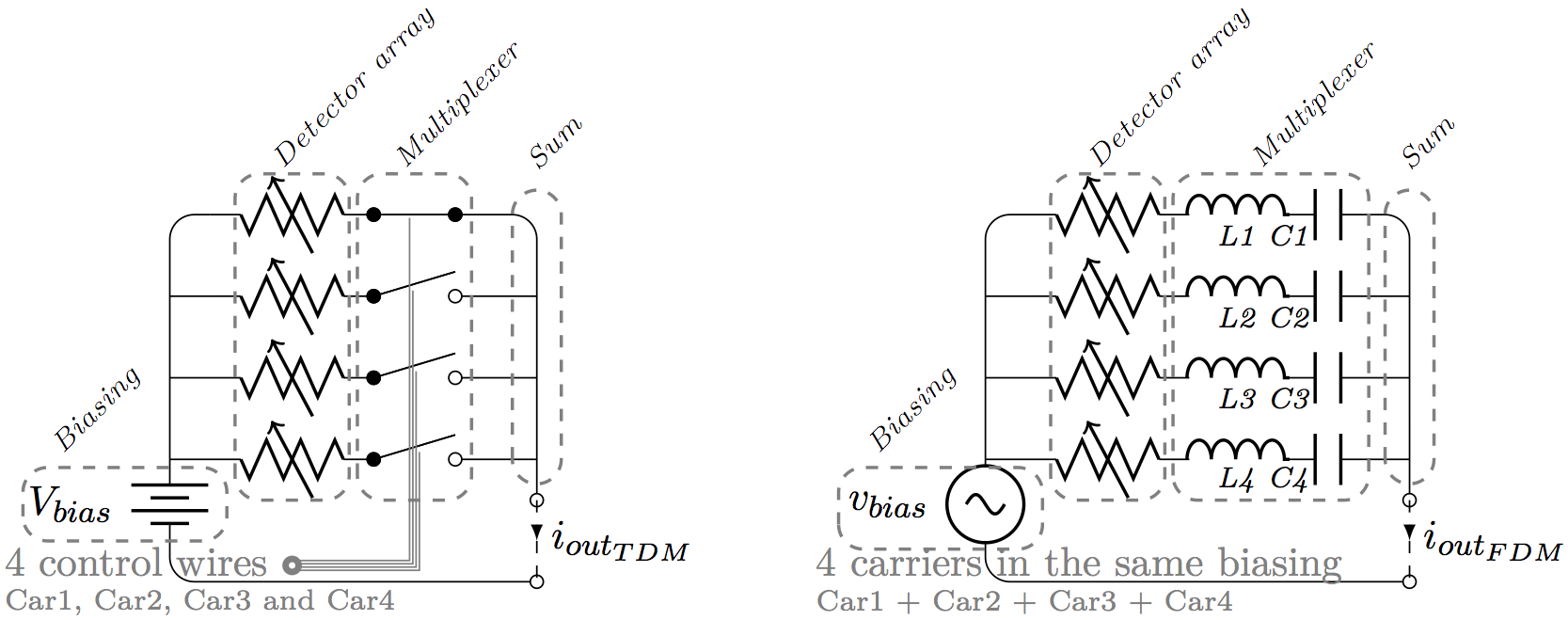}
	\end{center}
	\caption{Typical schematic of a TDM and FDM. Main part of the TDM is composed of switches. For FDM, $N$ different $LC$ filters are needed to {\it address} each sensors with different tones ($Car1$, $Car2$ ...).}
	\label{Fig13}
\end{figure}

Assuming again a 4 to 1 multiplexer example, figure \ref{Fig13} gives a typical front-end time domain and frequency domain multiplexer topology. Because we discuss the "front-end" multiplexing, sensors are part of the multiplexer itself. Considering sensors as variable resistors, a biasing is needed to convert resistance fluctuation to electrical signal. We chose, as an example, a voltage biasing (constant voltage applied to the sensors resistor) which implies a current readout represented by a simple wire where we measure $i_{out}$, the multiplexed electrical signal.

TDM need $N$ switches successively closed while FDM required $N$ LC filters with different resonance frequencies.

So, the main complexity of the TDM is the switch addressing system. The control wires used to address the switches are typically used for several multiplexer modules. This is a typical two diminution (2D) addressing: one diminution is used for addressing, the other for the readout. For instance, an array of $X$ lines and $Y$ columns could be divided in $Y$ sub-multiplexers, each following the topology of figure \ref{Fig13} left. The same $X$ control wires are then used to address sub-multiplexers. Doing that, $X+Y$ wires\footnote{$X$ control wires for addressing and $Y$ wires for readout the multiplexing signal of each sub-multiplexer.} are only required to readout $X \times Y$ sensors.

For FDM the complexity comes from the implementation of $N$ different capacitor values or inductors (or both) to select the carrier frequency associated to each sensor. Indeed, to avoid to use $N$ wires to bias $N$ sensors with their own carrier, only one wire is used with the $N$ carriers summed. So, an $LC$ filter allows to transmit only one of these carriers to the sensor. However, device inaccuracy leads to unavoidable errors on the resonance frequency of each resonator. This could reduce the separation between two carriers and thus lead to crosstalk between two sensors after demultiplexing. Moreover, the fact that $N$ different $LC$ filters are required implies that each sensor is not biased and read-out trough a similar circuit. Indeed, the small carrier frequency could be chosen of the order of the signal band-width $B$ (to fully use band-width from DC) and the frequency difference between 2 carriers is $2B$ at least. So, if we change, for example, only the capacitor values to change the resonance frequencies $\frac{1}{2\pi \sqrt{LC_X}}$, the $C_X$ value must go from a $C_0$ to a $C_{N-1} = (N-1)^2C_0$ capacitance for a $N$ to $1$ FDM. So, this implies huge parasitics and size disparities leading to different biasing and mismatch between sensors. In practice, the lower part of the spectrum is not used and the first resonance $f_0 = \frac{1}{2\pi \sqrt{LC_0}}$ is chosen far higher than $B$. However, even though we avoid values as high as $(N-1)^2C_0$, large variations of capacitance values could still be needed. Variation of both the $L$ and $C$ are sometimes needed to avoid too large (or too small) capacitors.

\subsection{QUBIC - the use of a time domain SQUID multiplexer}

The QU Bolometric Interferometer for Cosmology (QUBIC) is a cosmology experiment which aims to measure the B-mode polarization of the Cosmic Microwave Background \cite{bib2}. Its focal plane is mapped by 2 arrays of 1024 superconducting bolometers, each sensitive in the {\it mm} wave-length range. Operating at deep cryogenic temperatures (300~mK), bolometers \cite{bib3} must be multiplexed at very low temperature to reduce the number of read-out wires, and thus the thermal link to the room temperature ($\approx$~300K). This is achieved by using a time domain SQUID multiplexer.

\begin{figure}[h!]
	\begin{center}
		\includegraphics[height=.34\linewidth, keepaspectratio]{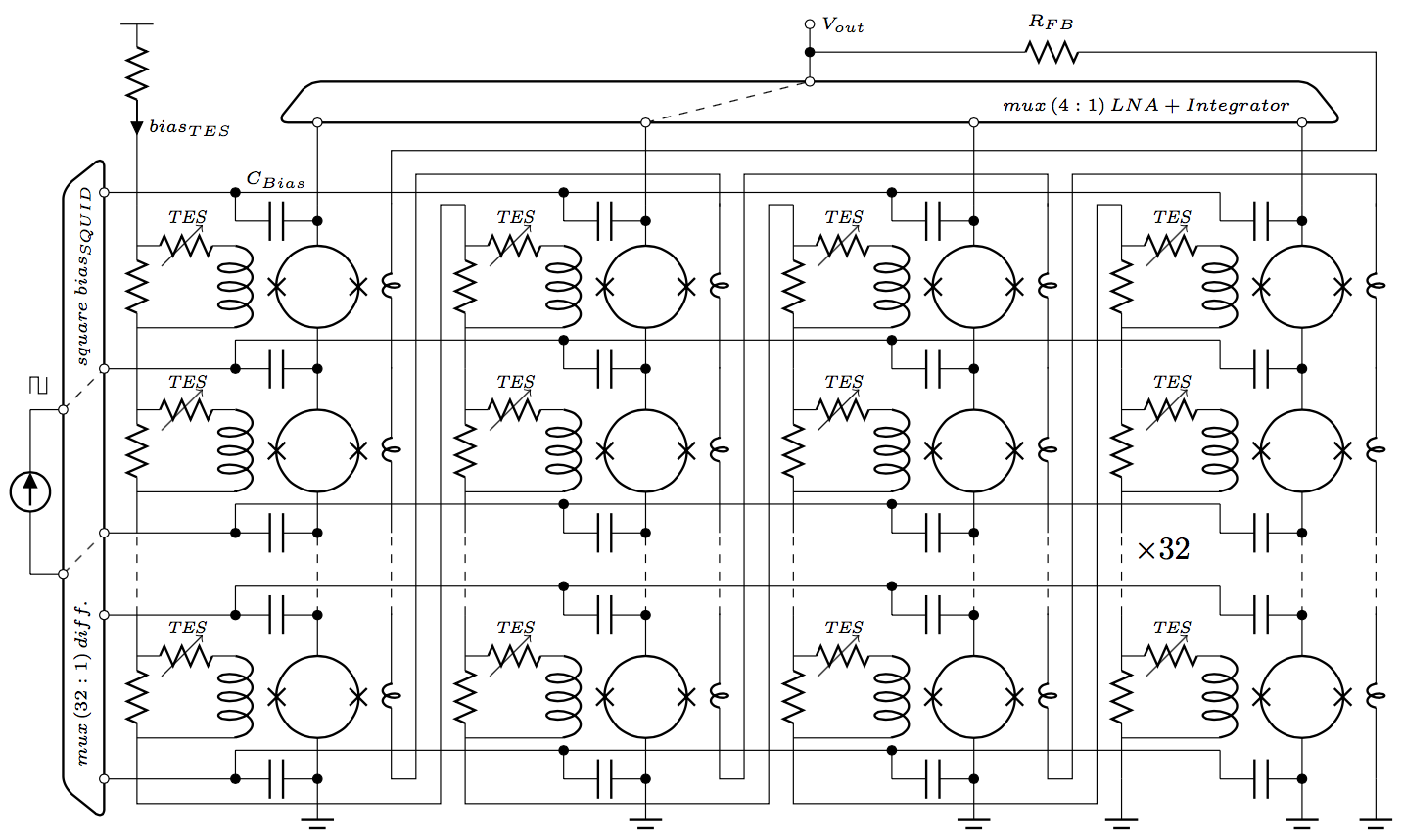}
		\includegraphics[height=.333\linewidth, keepaspectratio]{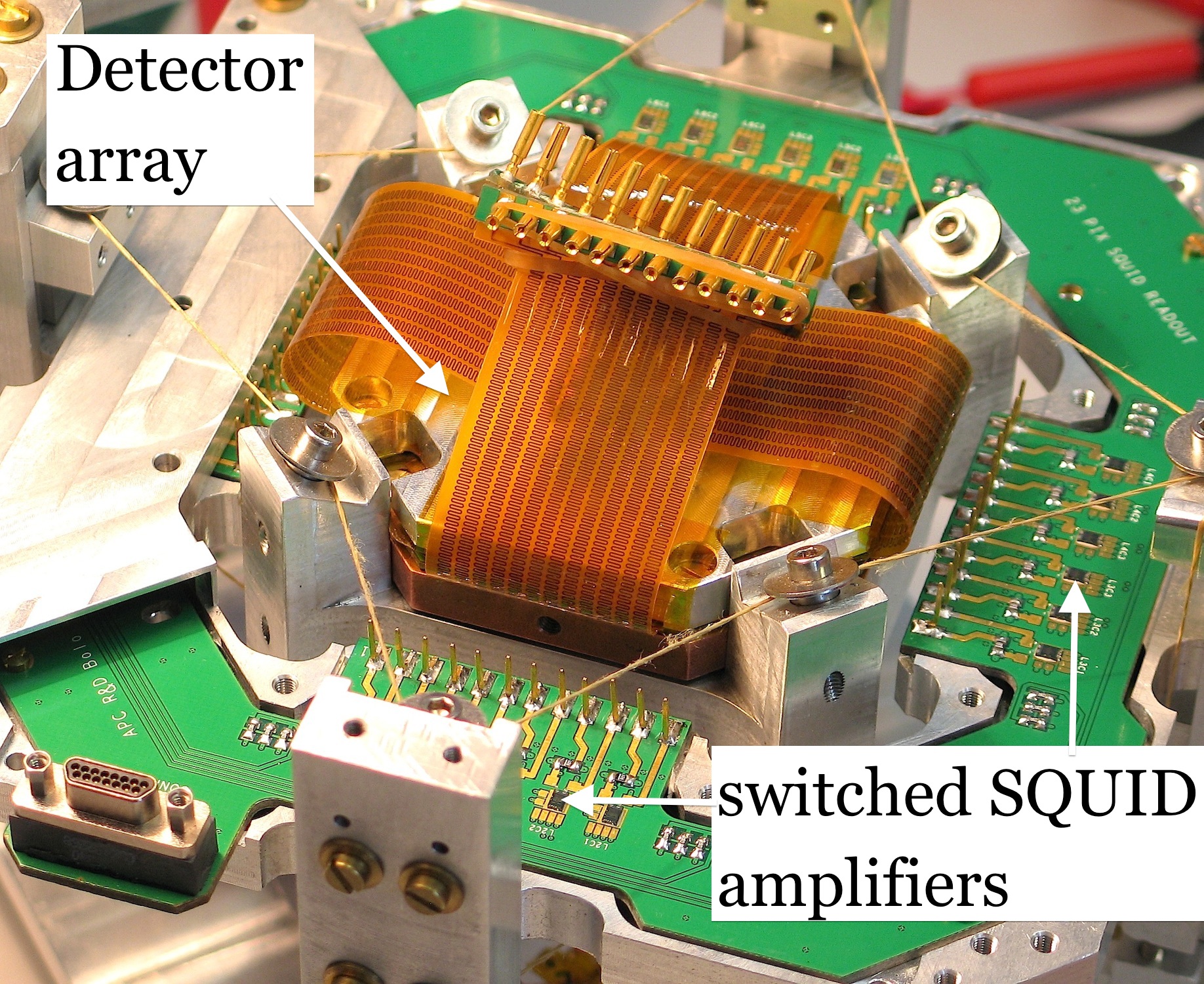}
		\includegraphics[width=1\linewidth, keepaspectratio]{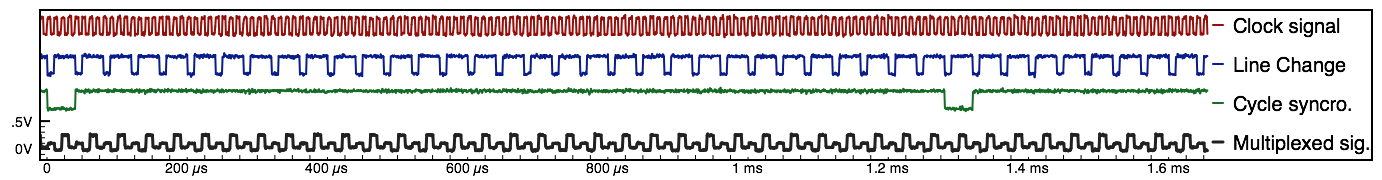}
	\end{center}
	\caption{Schematic, photography and wave-form (synchronization and multiplexed signals) of the time domain SQUID multiplexer developed for the QUBIC experiment \cite{bib5}.}
	\label{Fig14}
\end{figure}

Superconducting QUantum Interference Device (SQUID) are symbolized by a circle (superconducting {\it washer}), two crosses (Josephson junctions) and 2 loop inductors (current-to-flux conversion) on the schematic of the figure \ref{Fig14}. The larger loop inductor correspond to the input reading current coming from TES sensors (Transition Edge Sensors = superconducting bolometers). SQUID is used as the first amplification stage and could be turned ON and OFF by a cryogenic integrated circuit \cite{bib4} to play the role of the {\it switches}. The summation is obtained by reading the voltage across a column of SQUIDs. Indeed, sub-multiplexers are composed of SQUIDs connected together in series and biased sequentially. Figure \ref{Fig14} shows a 2D multiplexer composed by 4 sub-multiplexers (4 columns of SQUIDs). At the end, 4 multiplexed outputs (top part of the schematic in figure \ref{Fig14}) are readout sequentially by a cryogenic integrated circuit \cite{bib4, bib5}. The waveform of the multiplexed signal clearly shows the 4 common-mode voltages of the 4 SQUID columns. During the $1.28 ms$, the multiplexed signal transmits sequentially a sample of the 128 superconducting bolometer signals.

Photography in the figure \ref{Fig14} shows the SQUIDs (acting as switches) placed around a 24 TES array demonstrator. In addition to the sampling, the SQUIDs provide trans-impedance amplification allowing to read-out the TESs with room temperatures amplifiers outside the cryostat.

Other telescopes also dedicated to cosmic microwave background observation, in the same mm range, use frequency domain SQUID multiplexing to read-out bolometers \cite{bib6}, or kinetic inductance detectors (KID), which is based on a group of $LC$ resonators. $L$ is directly the sensitive part of the KID sensors and $C$ is chosen to tune the carrier frequency of each pixel thereby frequency multiplexed \cite{bib7}. However, we chose to describe a frequency domain SQUID multiplexer developed for high energy astrophysics in the {\it X-ray} range in the next sub-section.

\subsection{Athena X-IFU - the use of a frequency domain SQUID multiplexer}

The Advanced Telescope for High-ENergy Astrophysics (Athena) is an X-ray telescope designed to map hot gas structures in the Universe and to reveal black holes even in obscured environments \cite{bib8}. One of its instruments, the X-ray Integral Field Unit (X-IFU), is also based on a superconducting bolometer (TES) array coupled to X-ray absorbers. In this instrument, the TES is read-out by a frequency domain SQUID multiplexer.

\begin{figure}[h!]
	\begin{center}
		\includegraphics[height=.355\linewidth, keepaspectratio]{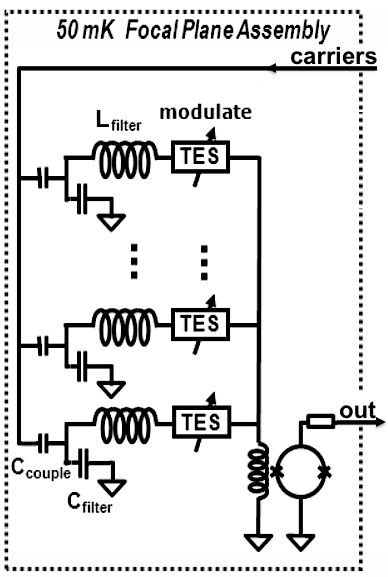} $\,$
		\includegraphics[height=.355\linewidth, keepaspectratio]{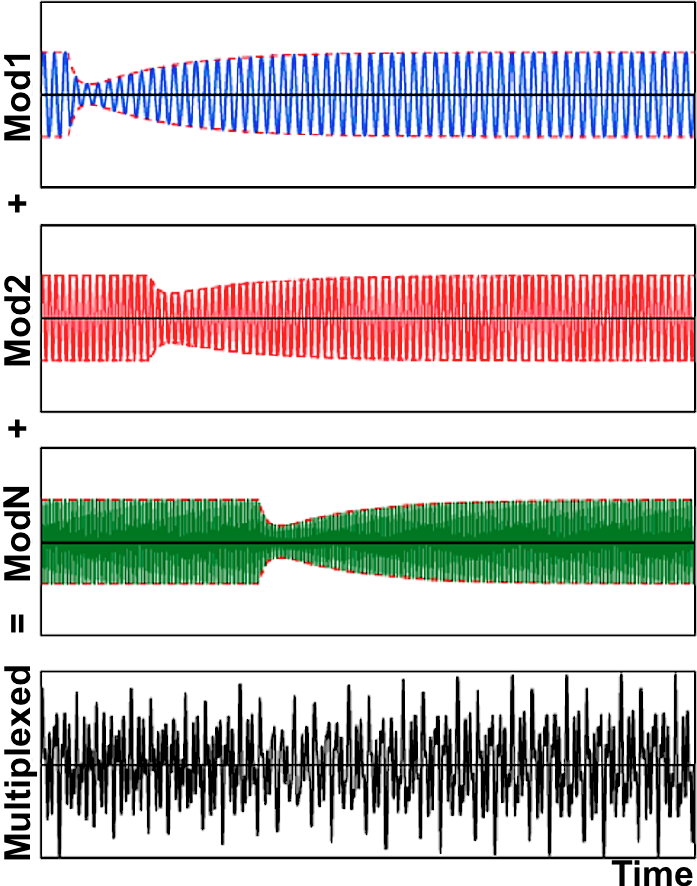}$\,$
		\includegraphics[height=.345\linewidth, keepaspectratio]{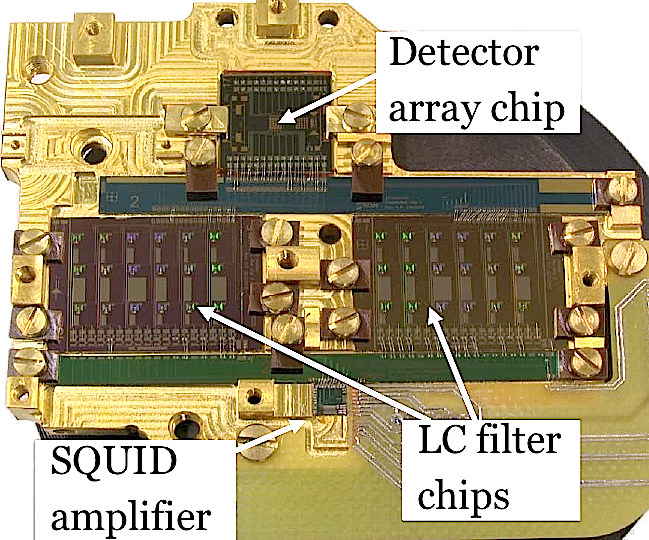}
	\end{center}
	\caption{Schematic, wave-form of the modulated/multiplexed signals and photography of a frequency domain SQUID multiplexer developed at SRON for X-ray astronomy \cite{bib9}.}
	\label{Fig15}
\end{figure}

Figure \ref{Fig15} shows a simplified schematic of one sub-multiplexer of the X-IFU instrument. $LC$ filters used to select the carrier frequency are clearly visible on the left of the schematic. Modulated signals are summed in current in the SQUID input loop. The TES AC bias is applied trough small capacitors ($C_{couple}$) to each resonator. This bias corresponds to the tones (comb frequencies) used to modulate each TES signal. Envelope of the modulated signals MOD1, 2 and 3 highlights the TES signal (a pulse in the case of the X-ray photon detection) modulating the carriers. The multiplexed signal is the summation of this modulated signal mixing the TES signal wave-forms.

Photography of the figure \ref{Fig15} shows the $LC$ filters close to the detector array with different capacitor sizes. This highlights the important role (and space) played by the $LC$ filters in such a frequency domain multiplexer. The detector array is in this case a small demonstrator array.

\section{Conclusion}

In this paper, we have discussed front-end multiplexing techniques, {\it i.e.} a simple way to reduce the number of wires needed to readout a detector array. The high channel capacities required to do multiplexing has been widely discussed. Finally, typical multiplexer topologies using switches or $LC$ filters has been discussed. Finally, two examples of cryogenic multiplexers using SQUID and developed for observation in astrophysics are described.
 
To conclude, we would like to insist on the fact that a multiplexing technique could be investigated only if the read-out chain has larger band-width and better noise or dynamic performances than that needed for one signal.

\acknowledgments

This review is based on the work on a cryogenic front-end multiplexer developed at APC in the {\it millimeter lab.} led by M. Piat, responsible of the QUBIC detection chain, and with the help and the main contribution of F. Voisin for cryogenic integrated circuit design and A. Tartari for his research on polarization sensitive Kinetic Inductance Detectors (KID). The author also thanks the SRON team involved in X-IFU instrument for fruitful discussion about FDM and baseband feedback.\\
These activities are supported by Centre National d'Etudes Spatiales (CNES) under the WFEE development for participation to the X-IFU instrument, Agence Nationale de la Recherche (ANR) under the QUBIC project, Paris Diderot University under the B-mode superconducting detectors (BSD) R\&D project and Centre National de la Recherche Scientifique (CNRS).

\end{document}